\documentclass[
   twocolumn,
   amsmath,amssymb,
   aps,
]{revtex4}

\usepackage{graphicx}
\usepackage{hyperref}

\usepackage{definition}

\begin{document}

\title{ Neutrino oscillations in the front 
        form of Hamiltonian dynamics}
\author{ Stanis{\l}aw D. G{\l}azek
and Arkadiusz P. Trawi\'nski }
\affiliation{ Institute of Theoretical Physics,
          Faculty of Physics, 
          University of Warsaw, Warsaw, Poland }
\date{26 August 2012 (submitted); 2 January 2013 (published)}


\begin{abstract}

Since future, precise theory of neutrino oscillations should 
include the understanding of the neutrino mass generation and a 
precise, relativistic description of hadrons, and observing 
that such a future theory may require Dirac's front form of 
Hamiltonian dynamics, we provide a preliminary front form 
description of neutrino oscillations using the 
Feynman--Gell-Mann--Levy version of an effective theory in 
which leptons interact directly with whole nucleons and pions, 
instead of with quarks via intermediate bosons. The interactions 
are treated in the lowest-order perturbative expansion in the 
coupling constants $G_F$ and $F_\pi$ in the effective theory,
including a perturbative solution of the coupled constraint
equations. Despite missing quarks and their binding mechanism, 
the effective Hamiltonian description is sufficiently precise 
for showing that the standard oscillation formula results from 
the interference of amplitudes with different neutrinos in 
virtual intermediate states. This holds provided that the 
inherent experimental uncertainties of preparing beams of 
incoming and measuring rates of production of outgoing particles 
are large enough for all of the different neutrino intermediate 
states to contribute as alternative virtual paths through which 
the long-baseline scattering process can manifest itself. 
The result that an approximate, effective front form theory 
reproduces the standard oscillation formula at the level of 
transition rates for currently
considered long-baseline experiments---even though the 
space-time development of scattering is traced differently 
and the relevant interaction Hamiltonians are constructed 
differently than in the commonly used instant form of
dynamics---has two implications. It shows that the common 
interpretation of experimental results is not the only one, 
and it opens the possibility of considering more precise 
theories taking advantage of the features of the front form 
that are not available in the instant form.
\end{abstract}

\maketitle

\section{ Introduction }

Contemporary theory of neutrino oscillations can be
developed using different forms of relativistic Hamiltonian
dynamics. \emph{A priori} possible forms were classified by
Dirac over 60 years ago~\cite{Dirac:1949cp}. He generally
distinguished three forms. In the first of these forms,
which is used most commonly and which Dirac called the
instant form (IF), a Hamiltonian $H$ generates the evolution
of a system in time $t$ of some inertial observer. This
means that the evolution is traced in terms of data that
change from one space-time hyperplane of constant $t$ to
another. These hyperplanes are called instants. In the
second form, called by Dirac the point form (PF), the
evolution is traced from one hyperboloid in space-time to
another. The PF distinguishes a point in space-time, and the
operators of spatial momentum in it, or generators of
translations in space, involve interactions. This is why 
the PF is not popular despite that the Lorentz symmetry is
represented in it purely kinematically.  In the third form, 
called by Dirac the front form (FF), the evolution of a 
system is traced from one space-time hyperplane of constant 
$x^+$ to another. In the conventional notation, $x^\pm = 
x^0 \pm x^3$, $x^\perp = (x^1, x^2)$, and all components of 
all tensors are handled in the same way. A hyperplane of 
constant $x^+$ is called a front. The FF evolution of a 
quantum system from one front to another is generated by 
the Hamiltonian $P^-$. 
     
Despite the existence of these options, neutrino oscillations
have until recently been described only using the IF of 
dynamics, e.g., in terms of the Feynman diagrams. In the 
Feynman diagrams, the IF is distinguished by using propagators 
ordered in time. Initially, Pontecorvo~\cite{Pontecorvo:1967fh} 
identified neutrino oscillations as a potential source of 
information on fundamental aspects of particle theory. 
Bilenky and Pontecorvo provided a quantum mechanical 
analysis~\cite{Bilenky:1977ne}, and Kayser~\cite{Kayser:1981ye} 
introduced wave packets for neutrinos. Rich~\cite{Rich:1993wu} 
constructed a space-time approach where neutrino emission, 
propagation, and absorption are treated as a single process. 
Giunti, Kim, Lee, and Lee~\cite{Giunti:1991ca,Giunti:1993se} 
described the oscillations in terms of the Feynman diagrams 
and wave packets. Grimus and Stockinger~\cite{Grimus:1996av}
considered neutron decay and antineutrino detection using 
electrons. A series of works followed~\cite{Beuthe:2001rc,
Giunti:2002xg,Giunti:2003ax,Cohen:2008qb,Akhmedov:2009rb,
Merle:2009re} and Akhmedov and Kopp~\cite{Akhmedov:2010ms} 
recently described the current status of the IF theory.
Major discussions of experimental results obtained using 
the IF interpretation of neutrino oscillations are reported 
in Refs.~\cite{Fukuda:1998mi,Ahmad:2002jz,Eguchi:2002dm,
Abe:2012gx,Adamson:2012rm}.

The formal scattering theory in the IF of dynamics has been 
recently applied to the neutrino oscillations using the 
approach of Gell-Mann and Goldberger~\cite{GellMann:1953zz}
with a slight extension~\cite{Glazek:2012pd}. An extension 
is needed because the formal theory in Ref.~\cite{GellMann:1953zz} 
assumes that the scattering region is small in comparison to 
the volume where the incoming particle beams are prepared and 
outgoing particles are detected while the neutrino oscillation 
experiments involve scattering regions that extend over so 
large a distance between particle sources and detectors that 
the scattering region in them is actually much greater than 
the entire particle acceleration and detection facilities. 
Thus, the neutrino oscillation experiments belong to the 
class of so-called long-baseline experiments which involve 
a long distance as an element in the measured observables. 

This article discusses neutrino oscillations using 
the FF of Hamiltonian dynamics in an effective 
theory~\cite{Feynman:1958ty,GellMann:1960np} as a
stepping stone to the required dynamical analysis 
in more fundamental theories. We distinguish three 
reasons for which a construction of the FF approach 
is of basic interest and why the main goal of this 
article is not a rederivation of the standard 
oscillation formula but to show that there exists 
a conceptual alternative to standard analysis and 
interpretation of experimental results that is
of value to particle theory.

The first reason we wish to distinguish is that the FF
offers a new relativistic interpretation of the neutrino
oscillation in terms of interference of amplitudes mediated
by virtual states of neutrinos. The new interpretation is
based on the operational definition of how one traces
evolution of quantum states that is different from that
based on the concept of simultaneity in the IF. Namely, one
uses the same basic principles that are used in the IF, but
the instants of some selected inertial observer are replaced
with the fronts. Using the FF of dynamics and counting $x^+$
instead of time $t$, one operates with results of
measurements that are correlated with suitably sent waves of
light. We shall discuss possible choices for how the fronts
can be defined and how the physical interpretation of
neutrino oscillations depends on these choices.

The second reason we wish to distinguish for introducing the
FF in description of neutrino oscillations is that the
vacuum problem~\cite{PhysRev.139.B684} in relativistic
quantum field theory is posed in the FF dynamics in a
different way than in the IF. The difference is a subject of
broad interest and scope and the relevant research on the
physics of the vacuum is too complex and too voluminous to
fully quote here~\cite{Nambu:1961tp,Kogut:1972di,Leutwyler:1977vy,
Shifman:1978bx,Gasser:1983yg,Weinberg:1988cp,Wilson:1994fk,
Brodsky:2009zd,Weinberg:2010wq,Brodsky:2012ku,PhysRevD.85.125018}. 
Consequently, we do not discuss the vacuum problem. Instead, 
we observe that the problem of a generation of masses of particles 
in the standard model~\cite{'tHooft:1972fi}, including the masses 
of neutrinos that are apparent in the neutrino oscillation,
can be associated with nontrivial properties of the vacuum.
Since the FF of Hamiltonian dynamics differs from the IF in
the approach to the physics of the vacuum, and thus may also
differ in its approach to searches for the dynamical origin
of the neutrino masses, one is motivated to ask if the FF 
could, in principle, be used to describe the neutrino oscillation 
at the current level of its understanding. This article provides 
a positive answer using the effective theory that is developed 
starting from the same effective Lagrangian 
density~\cite{Feynman:1958ty,GellMann:1960np} 
that was also used in Ref.~\cite{Glazek:2012pd} 
in the IF of Hamiltonian dynamics.  

The third reason we stress here is that the FF of
Hamiltonian dynamics is distinguished from the IF and PF by
the fact that 7 out of the 10 Poincar\'e group generators do
not depend on interactions in the FF, while in the IF and PF,
only 6 generators are free from interactions. The 7th
kinematical symmetry transformation is a boost along
the $z$ axis. Formally, the 7th symmetry implies that a hadron
structure appears the same to all observers related by a
boost along the $z$ axis. This class of observers includes the
observer at rest in a laboratory, with respect to whom a
hadron is at rest, and the observer in the infinite momentum
frame, for whom the same hadron moves practically with the
speed of light. Therefore, the FF of dynamics is considered
useful for the simultaneous theoretical explanation of hadron
structure in the context of spectroscopy, where the
constituent quark model guides phenomenology, and in the
context of high-energy scattering, such as in LHC, where the
parton model is used to describe the structure of hadrons.
Therefore, a complete, future theory of neutrino oscillations 
in which the coupling of massive neutrinos with leptons and 
quarks via massive gauge bosons will be fully understood is 
likely to require a FF Hamiltonian formulation. In this respect, 
it should also be mentioned that the FF of dynamics is useful 
in dealing with the issues of Fermi motion that will appear 
in experiments concerning neutrino oscillations at the theory 
level where leptons and quarks are treated on the same footing.

The article is organized as follows. Section~\ref{overview}
provides a brief overview of a typical long-baseline scattering
system that exhibits neutrino oscillations. We focus on the
example of an experimental setup resembling
T2K~\cite{Abe:2012gx}. Section~\ref{scatteringtheory}
recalls the Gell-Mann--Goldberger scattering theory with its
extension to the long-baseline experiments and FF of Hamiltonian
dynamics. Section~\ref{FFOF} describes key elements of the
calculation and interpretation of relevant amplitudes and 
transition rates. Section~\ref{sec:conclusion} concludes the 
article. Appendix~\ref{app:P-} describes the derivation of the FF
Hamiltonian used in the calculations described in Sec.~\ref{FFOF}. 

\section{ Overview of the scattering system }
\label{overview}

The neutrino oscillation can be observed in various experiments
that differ in terms of energies and momenta associated with
propagation of neutrinos. When neutrinos interact with
hadrons and the energies and momenta associated with their
propagation greatly exceed masses of nucleons, the
appropriate dynamical theory should involve quarks. However,
if the energies and momenta associated with neutrino
propagation are comparable with nucleon masses, one may
prefer to consider an effective theory in which neutrinos
couple to whole hadrons, instead of their constituents. 

The FF of Hamiltonian description of the neutrino oscillation
is developed in this article in the context of experiments 
resembling the T2K experiment~\cite{Abe:2012gx}, in which 
the neutrinos have energies and momenta on the order of 1 GeV. 
In this case, it is natural to follow Refs.~\cite{Feynman:1958ty,
GellMann:1960np} and assume that the appropriate effective
Lagrangian density for describing interactions between
neutrinos and protons, neutrons, pions, and muons is
\beq
\label{eq:cLI} \cL_I
& = &
   \frac{G_F}{\sqrt2}\cos\vartheta_C \;
   \bar \mu\gamma^\alpha(1-\gamma_5)\nu_\mu \;
   \bar p\gamma_\alpha(1-g_A\gamma_5)n
\nnn \\ &&
   -i\frac{F_\pi}{\sqrt2} \, 
   \bar \nu_\mu\gamma^\alpha(1-\gamma_5)\mu
   \; \partial_\alpha\pi^\dagger
   + H.c.\, .
\eeq
In this density, there appears the muon-neutrino field
$\nu_\mu$, which is a superposition of three fields of
neutrinos $\nu_i$ of different masses, $m_i$. Thus, $\nu_\mu
= \sum_{i=1}^3 U_{\mu i} \nu_i $.

In the FF Hamiltonian approach developed here, the evolution
of a quantum system is traced in the parameter $x^+ = x^0 +
x^3 \equiv t + z$ rather than $t$. In order to define the
time $t$ and the space coordinate $z$, one has to choose a
frame of reference. We first operationally define a fixed
frame of reference with coordinates $T$, $X$, $Y$, and $Z$ 
and then use it to describe our different choices of the 
space-time coordinates $t$, $x$, $y$, $z$ in the frames
in which the FF formalism is developed.

The long-baseline experiments are carried out on the Earth,
which rotates around its axis and circles around the Sun. 
Therefore, strictly speaking, a frame of reference with 
axes of fixed position with respect to the laboratories
such as Tokay and Kamioka does not define an inertial 
frame of reference that is suitable for developing any 
approach to neutrino oscillations using the concept of 
inertial observers. However, the corrections due to the 
Earth's nonuniform motion are small. 


The specific frame of reference with space-time coordinates 
$T$, $X$, $Y$, and $Z$ is fixed to the laboratories used in 
a long-baseline experiment. We shall call its coordinates the 
long-baseline coordinates. Let us focus attention on the example 
of T2K. In this case, the spatial origin of our long-baseline 
coordinate system is located in Tokay at the outlet of the 
neutrino source. The time $T$ is defined as the time of an 
observer at rest at this point. The $Z$ axis is directed to 
the muon detector in Kamioka $L \sim 300$ km away. This means 
that the $Z$ axis points about $2^\circ$ below the horizontal 
in Tokay in the direction of Kamioka. Let the $X$ axis be 
directed south and horizontally, and the $Y$ axis nearly 
vertically up, slightly tilted westward. The angular precision 
of determination of these axes directions corresponds to the 
ratio of the size of Kamioka detector (on the order of 30 m) 
to $2\pi L$, which is about $0.3'$.

\begin{figure}
\begin{center}
   \includegraphics[width=0.9\linewidth]{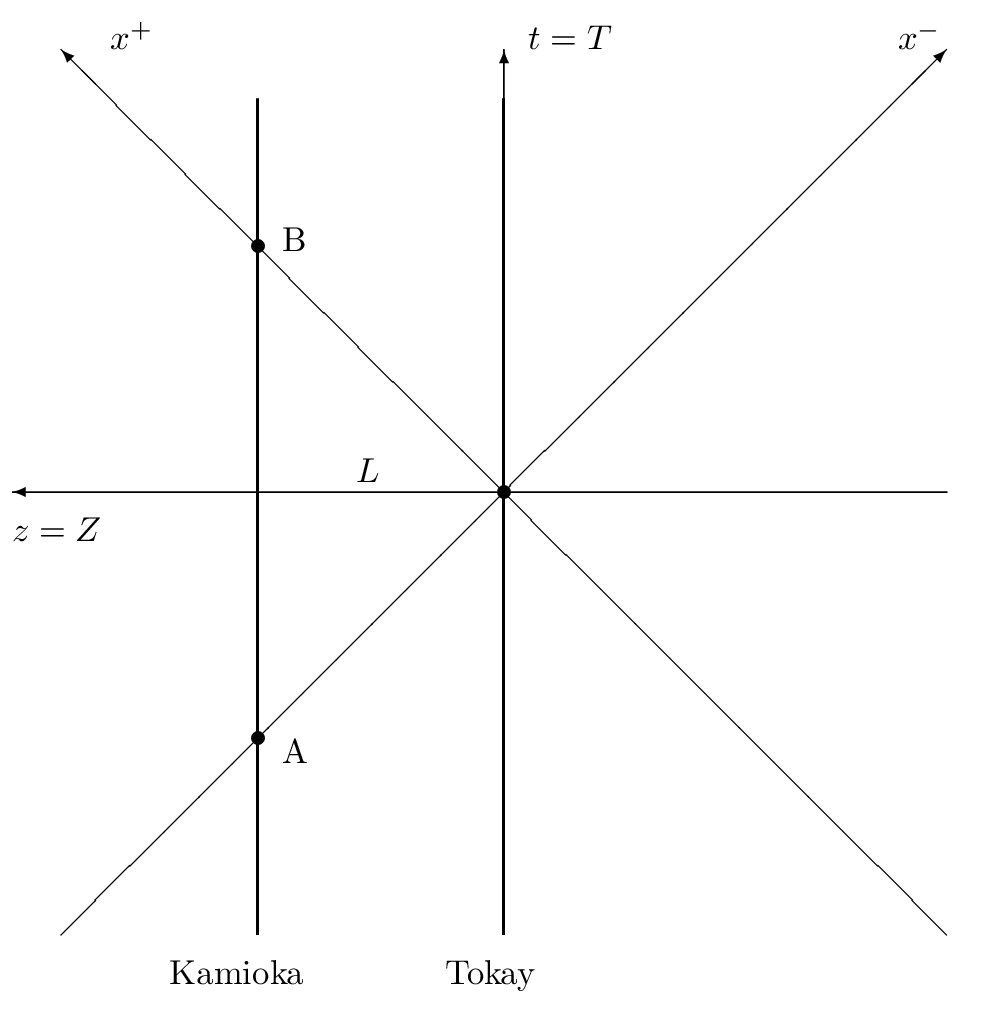}
   \caption[]{\label{referenceframesetup}
                   Example of the FF coordinates
                   defined using the long-baseline 
                   coordinates as $(t,x,y,z)=(T,X,Y,Z)$. 
                   This choice is called our {\it preferred} 
                   choice in the text.
                   The points $A$ and $B$ have coordinates
                   $x^+_A = 0$, $x^-_A = -2L$, $x^\perp_A = (0,0)$,
                   and
                   $x^+_B = 2L$, $x^-_B = 0$, $x^\perp_B = (0,0)$,
                   respectively, where $L$ is the long-baseline distance.
                   The thick vertical lines represent world lines of
                   laboratories located in Kamioka and Tokay. }
\end{center}
\end{figure}

The reference frame $(t,x,y,z)$ of an observer who
constructs the FF of Hamiltonian dynamics can be chosen in
many ways. We particularly distinguish the choice in which
$t=T$, $x=X$, $y=Y$ and $z=Z$, see Fig.~\ref{referenceframesetup}. 
We call this choice our {\it preferred} choice of the FF frame 
since it leads to a simple interpretation of the results 
obtained using the FF of dynamics. We shall discuss also 
other choices of $(t,x,y,z)$ with respect to $(T,X,Y,Z)$.

In the FF with our preferred choice of $z = Z$, the
scattering system evolves in $x^+$, which is measured along
the corresponding axis. This axis coincides with the world
line of a gedanken-experiment photon sent at time $T=0$ from the
long-baseline reference-frame origin in Tokay to Kamioka along
the $Z$ axis.

The evolution of the system in $x^+$ is generated by the 
FF Hamiltonian, denoted by $P^-$. Its interaction term 
corresponds to the Lagrangian space-time interaction 
density of Eq.~(\ref{eq:cLI}). The Hamiltonian is derived 
in Appendix~\ref{app:P-}, as an integral of the corresponding 
Hamiltonian density over the front $x^+=0$. The FF Hamiltonian
density is not as simply related to the Lagrangian density as
in the IF because of the constraint equations that are specific
to the FF. In the effective theory we use here, the constraint
equations, see Eqs.~(\ref{cnui}) to (\ref{cn}), are coupled 
and cannot be solved easily. Fortunately, in the case of
weak interactions, one can use an expansion in powers of the 
coupling constants $G_F$ and $F_\pi$. For our discussion of 
the neutrino oscillation, it is sufficient to solve the 
constraint equations using expansion up to terms of order 
$G_F F_\pi$. The resulting Hamiltonian contains all the terms
that are required for a description of the leading effect of 
neutrino oscillation and provide its FF interpretation. 
The generic interpretation involves the following experimental 
setup.

The $\pi^+$ beam is assumed to move nearly along the
$Z$ axis, a bit upward in the long-baseline coordinate system.
Let the $\bar\mu$ produced in decays of $\pi^+$ move at on even
greater angle upward. In such cases, the physical four-momentum 
transferred from Tokay to Kamioka via intermediate quantum
states, $p_\nu = (E_\nu, p_\nu^X, p_\nu^Y, p_\nu^Z)$, may,
in the majority of muon detection events in Kamioka, be 
reconstructed as lying close to the four-vector with components
$(E_\nu, 0,0, p_\nu^Z)$. This means that in our preferred
frame of reference, the physically likely four-momentum 
transfers have components close to the four-vector $p_\nu 
= (E_\nu, 0, 0, p_\nu^z)$ with $p_\nu^z = p^Z_\nu$. The 
corresponding FF momentum four-vector coordinates are 
$p^\pm_\nu = E_\nu \pm p_{\nu}^z = E_\nu \pm p_{\nu}^Z$ 
and $p^\perp_\nu = 0$. Of course, the FF components of $p_\nu$ 
would be different if the front $x^+=0$ were chosen in a 
different way. For example, if one reversed the $z$ axis, 
the FF components would be $p^\pm_\nu = E_\nu \mp p_\nu^Z$ 
and $p^\perp_\nu = 0$.

The FF formal scattering theory describes the preparation of
the $\pi^+$ beam in terms of its gradual buildup in $x^+$
rather than $t$. The beam-preparation time parameter $\tau^+$
in the FF of dynamics is introduced in a spirit of Gell-Mann
and Goldberger~\cite{GellMann:1953zz}. Using the convention
that $c=1$ and $\epsilon^- = 2/\tau^+$, the factor $e^{\cX^+/\tau^+} 
= e^{\epsilon^- \cX^+/2}$ is introduced to eliminate undesirable 
transients by the somewhat unphysical assumption
that a train of incident waves is released all at one time $\cX^+$.
Note that $\cX^+$ denotes the auxiliary FF ``time'' variable over
which one integrates in analogy to the integration over the
time variable $T$ in Eq. (2.3) in Ref.~\cite{GellMann:1953zz}. 

Physical interpretation of the pion beam preparation in the
FF of Hamiltonian dynamics depends on how one chooses the
$z$ axis in the long-baseline frame of reference. In our
preferred FF frame, the buildup of the initial pion and
neutron states has a relatively simple interpretation. The
pions emerge from a carbon target under ``pressure'' exerted
on it by highly energetic protons. The target has a shape of
a rod, its transverse cross-section diameter being much
smaller than its length. The uncertainty of created pion 
position (momentum) is comparable with the size (inverse of 
the size) of the carbon target, denoted by $r$ (from the word 
``rod''). With the use of electromagnets, pions produced at
various positions with various energies and moving at various 
angels are assumed focused to travel approximately in one 
direction. This increases the intensity of the pion beam and 
results in some energy distribution of pions in the beam. 
One could attempt to describe this distribution using a
carefully adjusted density matrix. One could also consider
issues of coherence, e.g., see Ref.~\cite{Akhmedov:2012uu}.
In this paper, we limit our consideration to the case of a 
monoenergetic pion beam with a well-defined momentum. 

We can estimate an upper limit on the value of a FF
beam-preparation time $\tau^+$ by considering the pion
lifetime. Namely, the length of the tunnel in which $\pi^+$
moves after being produced and in which it is nearly
certainly turned into $\bar \mu$ and $\nu_\mu$ by the
electroweak interactions is on the order of 100 m. Since the
pions travel with nearly the speed of light, they cover
approximately the same distance in time and in the $z$ direction
in our preferred frame. Thus, the estimate for the upper
limit on $\tau^+$ can be taken as 2 times the length of the
tunnel. This length is much shorter than the long-baseline
length $L$. Thus, the approximate, formal Gell-Mann and
Goldberger description of the pion wave function buildup
ought to contain a factor $ e^{\cX^+/\tau^+}$, where $\tau^+
\ll L$. Then, $\epsilon^- = 2/\tau^+$ is much greater than
$1/L$. The value of $\epsilon^-$ that corresponds to, at most,
100 m is on the order of at least $10^{-9}$ eV. At the same
time, the expected differences between the FF neutrino
``energies,'' $p^-_{\nu_i}$, with masses squared on the
order of $10^{-(3 \div 7)}$ eV$^2$ divided by $p_\nu^+ \sim
1$ GeV, are on the order of $10^{-(12\div 16)}$ eV, i.e.,
they are much smaller than $\epsilon^-$. This means that in
our preferred frame the FF energy eigenvalue density of
final eigenstates of $P^-$ has a width much larger than the
differences among eigenvalues of $P^-_0$ of virtual
neutrinos with different masses $m_i$ in the intermediate
states. This fact will be shown to lead to the validity of
the standard oscillation formula in the FF of Hamiltonian
dynamics. 

If we chose the $z$ axis in the opposite direction,
i.e., $z = -Z$ instead of $z=Z$, the carbon target, the pion
tunnel, and the proton and pion beams would all appear
nearly instantaneous in $x^+ = t + z$. However, the strength
of the pion wave function would not immediately be
normalized to 1 in the quantization volume according to the
Gell-Mann--Goldberger formulation of scattering theory.
Namely, following the Gell-Mann and Goldberger model for the
beam-buildup process in the IF, one can postulate that in
the corresponding FF model, the wave function strength should
smoothly increase with $x^+$ no matter how the $z$ axis is
chosen. In the case of $z = -Z$, one considers the incoming
plane-wave state $\ket{\phi_i}$ whose probability increases
with $\cX^+$ according to the function $e^{\epsilon^-
\cX^+/2}$ for $\cX^+ < 0$ for all values of $x^-$ and
$x^\perp$ ``simultaneously'' in the sense of $x^+$. 

Now, when $z = -Z$, the duration of the whole scattering
process in the sense of $x^+$ is also short. This happens
despite that the corresponding ``distance'' $x^-$ between
the pion source and final muon detector is very large, about
twice the distance $L$ between Tokay and Kamioka. In fact,
it will be shown below that $x^+ \sim (p^+_\nu/E_\nu)L$,
where $p^+_\nu$ is very small for $z=-Z$. In
Sec.~\ref{FFOF}, we will discuss the corresponding
quantum-mechanical mechanism by which the same oscillation
formula works irrespective of how the FF $z$ axis is
introduced. 

In the FF, we shall encounter instantaneous interaction
terms, in the sense of no delay in $x^+$. These interaction
terms are called seagulls~\cite{Brodsky:1973kb}. They correspond 
to the so-called $Z$ diagrams in the IF of dynamics in the 
infinite momentum frame, originating in virtual 
particles with momenta oriented opposite to the infinite
momentum. Since the meaning of simultaneity in the sense of
$x^+$ depends on how $x^+$ is defined, which in turn depends on the
choice of the $z$-axis, we will discuss the role of seagulls in
neutrino oscillations for different choices of the $z$ axis. The
same physical phenomenon of oscillation will result from
considerably different accounts of the flow of $x^+$ and
correspondingly different accounts of the intermediate
quantum states that depend on the different choices of
the $z$ axis.

\section{ Front form of scattering theory }
\label{scatteringtheory}

The theory objective in the case of experiments such 
as T2K is to calculate the transition rate from the 
state $\ket{\phi_i} = \ket{\pi^+ n}$ to the state 
$\ket{\phi_f} = \ket{p\mu\bar\mu}$ as a function of 
distance between the pion source and muon detector. 
The formalism does not depend in its essence on the 
details of initial and final states, and it is developed 
using the general notation of $\ket{\phi_i}$ and $\ket{\phi_f}$ 
for them, respectively. Effects of the Fermi motion of 
quarks in nucleons are ignored in the Feynman--Gell-Mann--Levy 
effective theory~\cite{Feynman:1958ty,GellMann:1960np}, 
and we neglect the Fermi motion of nucleons in nuclei.

According to the discussion in Sec.~\ref{overview}, the
incoming state $\ket{\Psi_i}$ is gradually built up
according to the formula
\begin{align}
&\ket{\Psi_i (x^+)}\nn
&\ = 
   {\epsilon^- \over 2} \int_{-\infty}^0 d\cX^+ \,e^{\epsilon^- \cX^+/2}
   \,e^{-iP^-(x^+-\cX^+)/2}\ket{\Phi_i(\cX^+)}\\
&\ = 
   e^{-iP^-x^+/2}\, \frac{i\,\epsilon^-\,}{p^-_i - P^-+i\,\epsilon^-}
   \ket{\phi_i}\, .
\end{align}
The state $\ket{\Phi_i(\cX^+)}$ is a function of $\cX^+$
that results from the action of $\exp{(-iP_0^-\cX^+/2)}$
on $\ket{\phi_i}$. The operator $P_0^-$ is the FF 
free Hamiltonian, such as in Eq.~(\ref{eq:P0}) 
in the case of T2K, and the state $\ket{\phi_i}$ is 
its eigenstate, with the corresponding eigenvalue 
denoted by $p_i^-$. 

The transition rate of the evolving system to the final
state $\ket{\phi_f}$ can be measured in terms of counting
particles in detectors (such as Kamiokande) using $x^+$ or
using $t$. It is natural for physicists to use $t$. This is
the time of an observer at rest in our long-baseline reference
frame (Kamiokande is at rest with respect to this frame).
Usage of $x^+$ appears less natural. However, the FF
counting of transition rates is physically quite realistic
in the sense of counting particles in coincidence with
the detection of light that defines the fronts of varying $x^+$.
In the long-baseline experiments, where the size of a neutrino
detector is negligible, one can take advantage of the fact
that the entire detector world line has a fixed value of
$z$. Thus, the measurement of a FF transition rate in terms
of states localized in the detector corresponds to the
differentiation of detection probability with respect to
$x^+$ and 
\beq
\label{dx+}
{\partial   \over \partial x^+} \es
{\partial t \over \partial x^+}\,{\partial \over \partial t}
+
{\partial z \over \partial x^+}\,{\partial \over \partial z} \, .
\eeq
For detectors located at fixed positions in the
long-baseline frame of reference, the operational
definition of transition rates in $x^+$ is
obtained by setting $\partial z/ \partial x^+=0$
and observing that $\partial t/\partial x^+=1/2$.

The probability that the system is in state 
$\ket{\phi_f}$ at $x^+$ is
\beq
 \omega_{fi}(x^+)
 \es  
 \frac{ |A(x^+)|^2}{||\Phi_f||^2 ||\Psi_i||^2} \, ,
\eeq
where the scattering amplitude calculated following
the Gell-Mann--Goldberger formulation of scattering 
theory is
\beq
 \label{eq:A(t)} 
 A(x^+)
 \es
 \bra{\phi_{f}}
 \frac{i\epsilon^- \, e^{i(p^-_f-P^-)x^+/2}}{p^-_i-P^-+i\epsilon^-}
 \ket{\phi_{i}}\, ,
\eeq
and the norms of states are not changing with $x^+$.
In the ratio of transition rates we are going to 
calculate, the norms cancel out, and they are omitted 
in further discussion (an alternative is to think 
about them as equal to 1). Using identity~(\ref{dx+}) 
for differentiation of $\omega_{fi}$ with respect to 
$t$, and omitting the norms, one obtains 
\beq
 {\partial \over \partial t}|A(x^+)|^2 \es
 {d \over dx^+/2} |A(x^+)|^2\, ,
\eeq
for detectors located at fixed $z$.
Following the steps analogous to the IF 
calculation~\cite{Glazek:2012pd}, one 
obtains the FF expression for the transition 
rate,
\beq
\label{eq:transition}
   \frac{d}{dx^+/2}\, |A(x^+)|^2
   & = &
   \frac{2\epsilon^-}{(p^-_f-p^-_i)^2 
+ (\epsilon^-)^2}|R^{\,\epsilon^-}_{fi}(x^+)|^2\, , 
\nn
\eeq
where $R^{\,\epsilon^-}_{fi}(x^+)$ is
\beq
\label{eq:R}
R^{\,\epsilon^-}_{fi}(x^+) & = &
   \bra{\phi_f} P_I^- e^{i(p_f^--P^-_0)x^+/2}
   \frac{i\epsilon^-}{p_i^- - P^- + i\epsilon^-}\ket{\phi_i}\,.
\nn
\eeq
This result is used in the next section
to derive the neutrino oscillation formula.

\section{ Front Form Oscillation Formula }
\label{FFOF}

There is a difference between the IF and FF calculations of
transition rates that results from a difference between the
Hamiltonians obtained from the same Lagrangian density of
Eq.~(\ref{eq:cLI}) in these two forms of dynamics. The
difference between the Hamiltonians is a consequence of the
constraints that appear in the FF and are absent in the IF 
of dynamics. The FF interaction Hamiltonian density is not
merely a negative of the Lagrangian interaction density
(with removed time derivatives of the pion field). Namely, 
it only depends on the dynamically independent components 
of fermion fields and contains terms that are instantaneous 
in $x^+$, called seagulls.

Appendix \ref{app:P-} describes the calculation of $P^-$.
The constraints in effective four-fermion theories, such as
in Eq.~(\ref{eq:cLI}), are a set of coupled nonlinear
equations. We can only solve them using a perturbative
expansion in powers of $G_F$ and $F_\pi$. Fortunately,
including terms order $G_F$, $F_\pi$, and $G_F F_\pi$ is
already sufficient for describing the dominant effect of
neutrino oscillations.

\subsection{ Calculation of $R^{\,\epsilon^-}_{fi}$ }

Denoting terms of order $G_F$ and $F_\pi$ as $P^-_1$, and terms
of second order including terms of order $G_F F_\pi$ as $P^-_2$,
one obtains from Eq.~(\ref{eq:R}) that
\begin{multline}
\label{RR}
R^{\,\epsilon^-}_{fi}(x^+)
= e^{i(p_f^--p_i^-)x^+/2} \\
\times \bra{\phi_f}   
   \left[
      P_1^-\frac{e^{i(p_i^--P_0^-)x^+/2}}{p_i^- - P_0^- + i\epsilon^-}P_1^-
     +P_2^-
   \right] 
\ket{\phi_i}\, ,
\end{multline}
where the phase factor in front is not important because of
the modulus in Eq.~(\ref{eq:transition}). The first term in
the square bracket describes transitions with a neutrino or
an antineutrino in the intermediate state. The second term
comes from the seagull interaction.

We denote by $p_\nu$ the four-momentum $p_\pi - p_{\bar\mu}$
that is physically transferred from $\pi^+$ to $n$. The
transfer is carried by a neutrino or antineutrino or mediated
by the seagull term. The four-momentum transfer does not
depend on the kind of neutrino that appears in the
intermediate state. The seagull term sums up effects coming
from constraints on neutrinos of all masses.

\emph{A priori}, there are two possible types of the intermediate
states, both coming in a sum over neutrino kinds labeled
using subscript $i$. For $p_\nu^+ > 0$, the intermediate
states contain $\bar \mu$, $n$, and $\nu_i$. For $p_\nu^+ <
0$, the intermediate states contain $\pi^+$, $p$, $\mu$, and
$\bar \nu_i$. No matter which possibility one considers for
incoming and outgoing particles, the ones for which the
reconstructed $p_\nu^+ > 0$ or the ones for which $p_\nu^+ 
< 0$, there is always an additional contribution from a 
seagull. The sign of reconstructed $p_\nu^+$ may \emph{a priori} 
also depend on the choice of the $z$ axis. However, when the 
experimentalists focus on events in which the reconstructed 
momentum transfer $p_\nu$ is obtained assuming that $p^+_\nu > 
0$, only states with virtual neutrinos can contribute, and 
antineutrinos are excluded. This is the case when the leading 
contributions to the total scattering amplitude come from the 
reconstructed four-momentum transfers $p_\nu$ that must be 
close to an on-mass-shell four-momentum of a neutrino with
some nonzero mass. These are the cases we focus on here. 
We shall come back to the possibility of studying antineutrino 
oscillation effects in Sec.~\ref{sec:otherz}.

In our preferred frame, every neutrino
carries a large positive $p^+_\nu$. The 
phase factor in the exchange term reads
\beq
 \exp\{i(p_i^--P_0^-)x^+/2\} \es
 \exp\{i(p_\nu^- -p_{\nu_i}^-)x^+/2\}\, . 
\nn
\eeq
Using the interaction Hamiltonian of 
Appendix~\ref{app:P-}, the matrix elements
that occur in Eq.~(\ref{RR}) are obtained in 
the forms
\begin{align}
\label{eq:matrices}
  &\bra{p\mu\bar\mu}P_1^-\frac1{p_i^- - P_0^- + i\epsilon} P_1^-\ket{n\pi^+}\nn
  & \ = -\,igf \, 2(2\pi)^3 \delta(p_i^+-p_f^+)\delta^{(2)}(p_i^\perp-p_f^\perp)\nn
  & \ \quad \times \bar u_p \gamma^\alpha(1-g_A\gamma^5)u_n\nn
  & \ \quad \times \sum_{j=1}^3 |U_{\mu j}|^2\bar u_\mu\gamma_\alpha(1-\gamma^5)
   {1\over p^+_\nu}\frac{\slash p_{\nu_j} + m_{\nu_j}}
   {p_\nu^- - p_{\nu_i}^- + i\epsilon^-}\nn
  & \ \quad \times (-ip_{\pi}^\beta) \gamma_\beta(1-\gamma^5) v_{\bar\mu}\, , \\
\label{eq:P2}
  &\bra{p\mu\bar\mu} P_{gf}^- \ket{n\pi^+}\nn
  &\ =  -\,igf \,  2(2\pi)^3 \delta(p_i^+-p_f^+)\delta^{(2)}(p_i^\perp-p_f^\perp)\nn
  &\ \quad\times \bar u_p \gamma^\alpha(1-g_A\gamma^5)u_n\nn
  &\ \quad\times\,\bar u_\mu \gamma_\alpha(1-\gamma^5)
   \frac{\gamma^+}{2 p^+_\nu}(-ip_{\pi}^\beta)
   \gamma_\beta(1-\gamma^5) v_{\bar\mu}\, ,
\end{align}
where $P^-_{gf}$ obtained from the density of 
Eq.~(\ref{Pgf}) is the only term in $P_2^-$ 
that contributes.

\subsection{ Neutrino exchange vs seagull }

Our goal now is to show that the oscillating exchange term
in Eq. (\ref{RR}) dominates in the transition rate. 

The two contributions in the square bracket in Eq.
(\ref{RR}), shown in Eqs. (\ref{eq:matrices}) and
(\ref{eq:P2}), differ only by the following factors that
appear between the two current factors that are expressed in
terms of the spinor matrix elements associated with the
pion-decay and neutron-proton transition:
\beq
 \frac1{p^+_\nu}\frac{\slash p_{\nu_i} + m_{\nu_i}}
 {p_\nu^- - p_{\nu_i}^- + i\epsilon^-} 
 & {\rm vs} &     
 \frac{\gamma^+}{2 p^+_\nu} \, .
\eeq
These two terms can be looked at as contributing
$\gamma$ matrices with various coefficients. The seagull
contributes only to the coefficient of $\gamma^+$. One can
see the relative size of the two terms by first factoring
out the Feynman-like denominator,
\beq
   {1 \over D_i(p_\nu)} \es
   {1 \over p^+_\nu (p_\nu^- - p_{\nu_i}^- + i\epsilon^-)} 
   =  
   {1 \over p_\nu^2 - m_{\nu_i}^2 + i\epsilon^-p_\nu^+ }\,,
   \nnn\\ 
\eeq
in both of them, and then comparing (assuming $p_{\nu_i}^\perp=0$) 
\beq
  p_{\nu_i}^+ {\gamma^- \over 2} 
  +
  p_{\nu_i}^- {\gamma^+\over 2} 
  + m_{\nu_i} 
  ~~{\rm vs}~~
  (p_\nu^- - p_{\nu_i}^- + i\epsilon^-p_\nu^+) {\gamma^+\over2}.
\eeq
The second term, which comes from the seagull, is as small
as the off-shellness of the neutrino, and the first term contains
$p_{\nu_i}^+$ and $p_{\nu_i}^-$, of which at least one is
large. Whether $p_{\nu_i}^+$ or $p_{\nu_i}^-$
is large depends on the choice of $z$ axis. The
FF smallness of the seagull term in comparison with the 
neutrino exchange term corresponds to the IF smallness of 
the antineutrino exchange terms in comparison with the 
neutrino exchange terms in the Hamiltonian description 
of neutrino oscillations. One should expect this result 
since the seagull interaction terms in the FF Hamiltonians 
correspond to the terms in the IF perturbation theory in 
which an intermediate particle carries a negative fraction 
of the system total momentum in the infinite momentum
frame---see Sec. V of Ref.~\cite{Brodsky:1973kb} and 
Refs.~\cite{Kogut:1969xa,Bjorken:1970ah,Neville:1971zk,
Neville:1971uc,Chang:1972xt,Chang:1973qi,Yan:1973qf,
Yan:1973qg,TenEyck:1974cy}---in this case the relevant 
particle being the antineutrino.

Thus, the exchange term with an oscillating phase factor is
expected to dominate the transition rate. We have inspected
matrix elements of the above matrices in randomly selected
momentum and spin configurations that may be relevant in
experiments such as T2K. In all cases we so sampled, our
inspection confirmed that the above-described analysis of
the coefficients of $\gamma$ matrices correctly estimates
the actual ratio of the full amplitudes, and the seagull
contributions can be neglected in comparison with the
exchange term contributions. Once the seagull term is
neglected, the sum over intermediate states with different
neutrinos with appropriate phases is inserted in
Eq.~(\ref{eq:transition}).

\subsection{ FF interpretation of the oscillation }

In a theory of the quantum mechanical scattering process,
one can fix momenta of the initial and final particles.
These momenta define the momentum transfer four-vector
$p_\nu$. The physical transfer of $p_\nu^-$ can greatly
differ from $p_{\nu_i}^-$, since the latter is calculated
from the neutrino-mass-shell condition for given $p^+_\nu$
and $p^\perp_\nu$, while the former results from the
difference of momenta of scattering particles. However, due
to the denominator $D_i(p_\nu)$, the scattering amplitude is
large only for $p_\nu^-$ near one of the values
$p_{\nu_i}^-$. In experiments such as T2K, it may be assumed
that the observed counts of muons in the far detector come
from these large terms. Since the actual size of the
scattering amplitude also depends on the uncertainty of the
FF energy $\epsilon^-$, the final interference pattern
between the amplitudes coming from specific intermediate
states also depends on the size of $\epsilon^-$.

\begin{figure}[!ht]
   \includegraphics[width=0.95\linewidth]{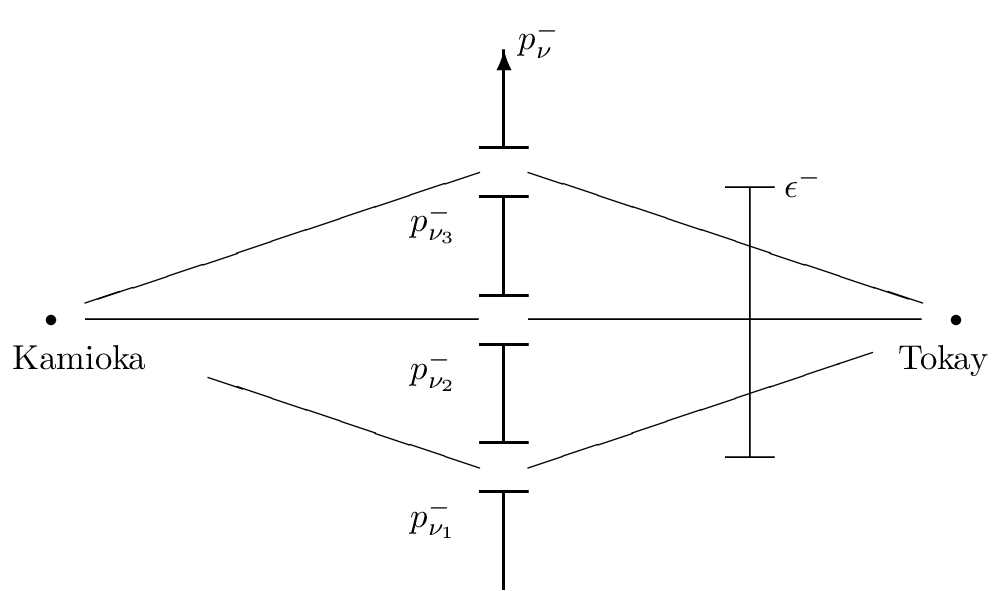}
   \caption[]
   {\label{fig:yang}
     Schematic representation of the interference of amplitudes
     mediated by virtual states with neutrinos of different FF free 
     energies $p^-_{\nu_i}$, due to the different masses $m_i$. The
     $p_{\nu_i}$ increase with the mass $m_i$; $p_{\nu_1} \leq p_{\nu_2} 
     \leq p_{\nu_3}$. Like in the elementary quantum slit experiment, 
     amplitudes with certain values of $p_\nu^-$ provide dominant 
     contributions to the interference pattern. The FF energy 
     uncertainty $\epsilon^-$ must be large enough to create the 
     standard interference pattern in the total counting rate of 
     muons in the far detector.}
\end{figure}

Figure~\ref{fig:yang} shows that the three amplitudes that
contribute to counting muons coming from interactions that
involve three different intermediate states with
different-mass virtual neutrinos can interfere in analogy
with the interference pattern that is familiar from the
elementary quantum slit interference experiment, except that
the role of the slit position is played by the FF on-mass-shell
energy $p^-_{\nu_i}$. The number of ``energy slits''
available for the interference depends on the size of
$\epsilon^-$, which must be large enough to enable the
interference of all potentially available states with 
virtual neutrinos.

\begin{figure}[!ht]
   a)
   \includegraphics[width=0.9\linewidth]{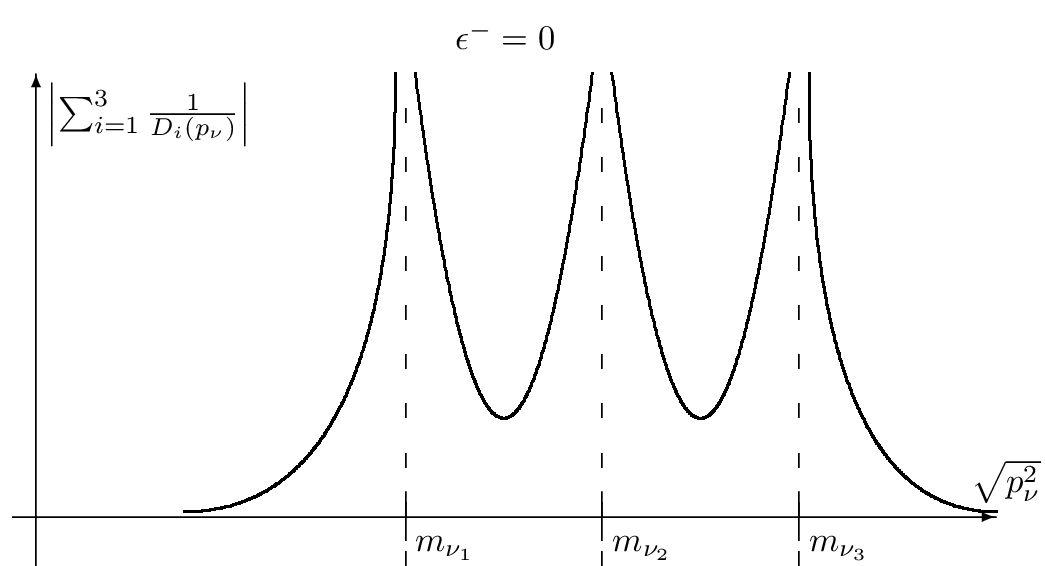}
   \vskip2em
   b)
   \includegraphics[width=0.9\linewidth]{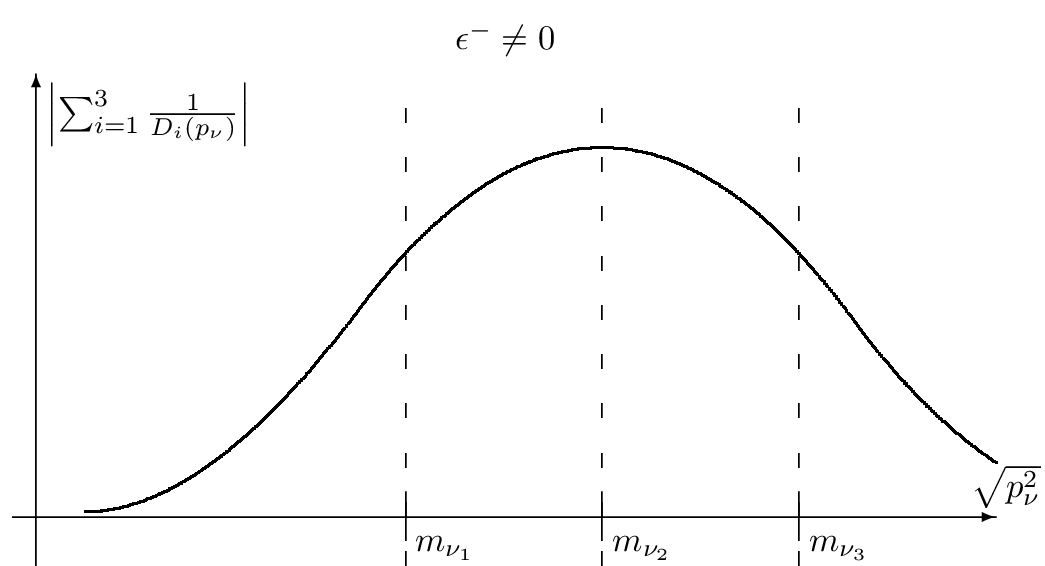}
   \caption[]
   {\label{fig:epsilon}
  Qualitative plots of the function $|\sum_{i=1}^3 1 /
  D_i(p_\nu)|$ using arbitrary units for a) $\epsilon^- = 0$,
  and b) $\epsilon^- \neq 0$. The neutrino masses $m_{\nu_i}$
  are assumed to increase with $i$; $m_{\nu_1} \leq m_{\nu_2}
  \leq m_{\nu_3}$. The scales are arbitrarily adjusted for the
  purpose of obtaining a clear drawing. The larger $\epsilon^-$
  in comparison to $\Delta m^2_{ij}L/p^+_\nu$, the more accurate
  the standard oscillation formula, because the wider the width
  $\epsilon^- p_\nu^+$, the more neutrino intermediate states uniformly
  contribute to the muon counting rate in the far detector.}
\end{figure}

For $x^+ = 2L$, the neutrino mass-dependent phase factors present in
the complete scattering matrix element $R^{\,\epsilon^-}_{fi}(x^+)$ 
in Eq.~(\ref{eq:transition}) can be reduced to simpler forms,
$ \exp(i p_{\nu_i}^- 2L / 2) \rightarrow \exp( i m_{\nu_i}^2 L /  
p_\nu^+)$, since the transverse and $+$ momenta are the same for 
all kinds of the contributing neutrinos, and the common phase factor 
due to $p_\nu^\perp$ is irrelevant to the interference pattern. 
Thus, the amplitude $R^{\,\epsilon^-}_{fi}(x^+)$ in Eq.~(\ref{eq:matrices}) 
significantly varies with $p_\nu$ only due to the identified above mass 
dependent phase factors and FF energy denominators for different 
neutrinos. The result for $R^{\,\epsilon^-}_{fi}(2L)$ is proportional to   
\beq
   \sum_i |U_{\mu i}|^2{e^{i\,m_{\nu_i}^2 L /  p_\nu^+}\over D_i(p_\nu)}
   \es
   \sum_i |U_{\mu i}|^2\frac{e^{i\,m_{\nu_i}^2 L /  p_\nu^+}}
   {p_\nu^2 - m_{\nu_i}^2 + i\epsilon^-p_\nu^+ }\,.\nnn\\
\eeq
The statement made above that the interference of amplitudes with different 
neutrinos in the intermediate states occurs for a sufficiently large value of 
the uncertainty $\epsilon^-$, is illustrated in Fig.~\ref{fig:epsilon}.
When the size of $\epsilon^-$ exceeds the differences among the 
on-mass-shell FF energies $p^-_{\nu_i}$, all neutrino
``channels'' or ``slits'' in $p^-_\nu$ contribute, and the
corresponding total amplitude exhibits the standard
interference pattern.

On the other hand, a reduction of the ratio of the
experimental $p^-_\nu$ uncertainty, $\epsilon^-$, to the FF
energy differences among different neutrinos would lead
to deviations from the standard oscillation formula, since
some channels or slits in $p^-_\nu$ would no longer
be able to contribute to the lepton counting in the far
detector. For example, heavy sterile neutrinos with
sufficiently large masses must drop out form the interference
pattern with relatively small uncertainty $\epsilon^-$.

It is clear that the experimental and theoretical ways of
studying deviations from the standard oscillation formula
require further investigation. Namely, the patterns of
deviation depend on the neutrino masses and coupling constants,
including the factors $U_{\mu i}$. Therefore, the deviations
are sources of information on these parameters. The way to
reduce the relevant ratio or, equivalently, to enhance the
FF energy differences in comparison to $\epsilon^-$
using the front choice of $z=Z$ is to reduce $p_\nu^+$. In
order to have a possibility of reducing $p^+_\nu$ and still
creating the final lepton, one might consider detection of
electrons instead of muons.

\subsection{ Other choices of $z$ axis }
\label{sec:otherz}

In the experimental T2K-like setups where the observed 
neutrino interference patterns are unambiguously identified 
using reconstructed four-momentum transfers $p_\nu$, one
can establish if the physical $p_\nu$ must approximately 
match the value that corresponds to a $\pi^+$ decay
into some free neutrino and $\bar \mu$. If it does, 
$\nu_i$ and $\bar \mu$ may only form states with a free 
invariant mass comparable to the $\pi^+$ mass and with a 
relatively small uncertainty determined by the product 
$\epsilon^- p^+_\nu$. In such cases, the $p^+_\nu$ is always 
greater than 0 no matter how one chooses the $z$ axis to 
define the front. All intermediate states of the virtual 
particles that can significantly contribute contain only 
neutrinos. The seagull contribution is always small.

Still, the theoretical FF interpretation of the oscillation
changes when one changes the $z$ axis. A particularly
instructive example of a change is provided by the case of
$z = -Z$; see Sec.~\ref{overview}. In this case, assuming
$p^\perp_\nu \sim 0$, the FF time interval $x^+$ during
which the neutrino exchange occurs is estimated as the time
difference between the duration of a flight of a real
massive neutrino and a massless photon from Tokay to
Kamioka, approximately equal to $(p^+_\nu/E_\nu) L$. For a
neutrino on the mass shell $m_i$, one obtains $x^+ \sim
x^+_i \sim (m_i^2/2 E_\nu^2)L$, instead of $2L$ obtained in
the case $z=Z$ discussed in the previous section. But the
corresponding neutrino's FF energy is now approximately
$p_{\nu_i}^- = 2 E_\nu$. Therefore, the half of a product 
of a short $x^+_i$ and a large $p_{\nu_i}^-$ that appears 
as a phase in the exponent is the same as the result $m_i^2 L/
(2E_\nu)$ in the case $z = Z$. The width of Feynman-like 
denominators is also obtained without change. Namely,
although the $p_\nu^+$ for $z = -Z$ is very small, the
pion beam is prepared nearly instantaneously in $x^+$
in comparison to $2L$. This means that $\epsilon^-$ is 
large and the product $\epsilon^- p^+_\nu$ is not changed. 
In summary, the intermediate neutrinos that are nearly 
instantaneously exchanged in $x^+$ for $z=-Z$ have very 
small $p^+_\nu$ but a huge $p^-_{\nu_i}$, and hence the 
oscillation pattern is not expected to change by changing 
the front despite that the FF interpretation depends on the
choice of a front. Thus, although the FF interpretations 
differ among themselves and from the IF interpretation that 
uses the laboratory time as the parameter of evolution, the 
approximate oscillation formula is obtained in the same form.

Besides the two choices of $z = Z$ and $z = -Z$, there
exists a whole set of choices with the $z$ axis directed 
at some angles to the $Z$ axis. In principle, one can 
also discuss reference frames in motion with respect 
to the long-baseline frame. Instead of discussing these 
multiple options, we prefer to point out a qualitatively 
different aspect of the FF interpretation of the neutrino 
oscillation.

Namely, one may ask if the four-momentum transfer $p_\nu$
that is reconstructed in analyzing observed events must
always be close to a $p_\nu$ that corresponds to a real
decay of $\pi^+$ or, perhaps, it could involve $p^+_\nu <
0$. In the latter case, the $x^+$-dependent phase factor for
$z = -Z$ could be interpreted as due to a free evolution of
the antineutrino that nearly instantaneously in $x^+$
travels from Kamioka to Tokay, where it is absorbed by a
pion. The authors have not succeeded in finding out if
even in principle the existing or prospective data on the 
neutrino oscillation permit a reconstruction of $p_\nu$ 
without assuming that it must approximately correspond to 
a real pion decay, instead of absorption of an antineutrino 
that nearly instantaneously in $x^+$ is created in Kamioka 
and subsequently absorbed in Tokay. It is also not clear if 
such reconstruction, even if possible, could be associated 
with a measurable counting rate of muons in Kamioka. The only 
observation that we can offer is that the invariant mass
difference between a neutron and a proton-muon pair implies
that the reconstructed $p_\nu$ would have to correspond to 
an antineutrino virtuality on the order of 100 MeV,
which suggests greatly reduced muon counting rates.

On the other hand, the intriguing aspect of considering the
reconstructed $p^+_\nu$ close to 0 is that the FF vacuum
problem can be associated with the region of $p^+_\nu = 0 
$~\cite{Kogut:1972di,Wilson:1994fk,Brodsky:1997de}. Thus,
the opportunity for reconstructing momentum transfers with
extremely small $p^+_\nu$ in long-baseline neutrino oscillation
experiments could perhaps be used to test theoretical ideas
concerning vacuum involvement in the generation of neutrino
masses. Studies of the small $p^+_\nu$ region would have to 
include a precise description of the Fermi motion effects 
associated with the binding of a neutron in the oxygen nucleus 
in the far detector (in the case of water detectors) and with 
the binding of quarks inside hadrons. In any case, the region 
of $p^+_\nu = 0$ is singular in the quantum field theories 
that form the standard model and deserves a formal study in 
the context of neutrino oscillations.

\section{ Conclusion }
\label{sec:conclusion}
   
The FF of dynamics provides an alternative theoretical 
description of neutrino oscillations to the descriptions 
available in the IF.
Since the FF also provides an alternative formulation 
of the ground-state problem in quantum field theory, 
including the concepts of vacuum condensates and mass 
generation, we conclude that the neutrino oscillation 
can be studied using the FF of dynamics in conjunction 
with the fundamental issues of particle theory.

The FF approach provides an interpretation of the standard
neutrino oscillation formula as resulting from the
interference pattern that occurs only when the experimental
front form energy uncertainty $\epsilon^-$ is sufficiently large 
in comparison to the differences between the individual
on-mass-shell values of $p^-_{\nu_i}$ for any of the
neutrinos of the kind $i$. The FF explanation of the 
interference leads to the condition that when the differences 
between individual $p^-_{\nu_i}$, for the same $p^+_\nu$ 
and $p^\perp_\nu$, are much greater than the uncertainty 
$\epsilon^-$, the standard oscillation formula is not valid. 
While the conditions of validity of the standard oscillation
formula are well-satisfied in the case of experiments like
T2K and 3 already known neutrinos with quite small masses, 
they are not satisfied for much heavier neutrinos.

\begin{acknowledgments}
This work was supported by the Foundation for Polish Science
International Ph.D Projects Programme co-financed by the EU
European Regional Development Fund.
\end{acknowledgments}

\appendix

\section{\label{app:P-} Calculation of $P^-$}

The operator $P^-$ is a generator of translations in $x^+$.
It is defined in the FF of dynamics through an integral 
over $x^-$ and $x^\perp$,
\beq
   P^- & = & {1 \over 2} \int dx^- d^2x^\perp \,T^{+-}\,,
\eeq
where $T^{\mu \nu}$ is the energy-momentum density tensor,
and half of $T^{+-}$ will be denoted by $\cP^-$.

Standard methods of canonical quantization~\cite{Chang:1972xt} 
lead to the expression for $P^-$ that corresponds to the 
Lagrangian density $\cL = \cL_0 +\cL_I$, where
\beq
\cL_0
   &=& 
   \partial_\mu \pi^\dagger \partial^\mu \pi -
   m_\pi^2 \pi^\dagger \pi
   + \sum_{\psi} 
   \bar \psi(i \slash\partial - m_\psi  ) \psi \, ,
\eeq
and $\cL_I$ is given in Eq.~(\ref{eq:cLI}),
according to the formula
\beq
   \label{eq:Tmunu}
   T^{\mu \nu} \es g^{\mu \alpha} 
   \sum_\phi {\partial \cL \over \partial
   \partial^\alpha \phi } \partial^\nu \phi 
   - g^{\mu \nu} \cL \, .
\eeq
Summation over $\phi$ is meant to indicate that
one sums over all fields in the theory. $T^{+-}$ 
is expressed in terms of the fields and their 
conjugated momenta in order to quantize the 
theory.

The canonical conjugate momenta for the fermion fields,
$\partial \cL / \partial \partial^- \psi = \bar \psi
i\gamma^+/2$, depend on the field components $\psi^{(+)} 
= \frac12\gamma^0\gamma^+\psi$, which are the dynamically 
independent variables. Matrices $\Lambda^\pm = \frac12\gamma^0 \gamma^\pm$ 
are projectors. The fermion field components $\psi^{(-)} 
= \frac12 \gamma^0 \gamma^-\psi$ satisfy constraint 
equations which couple all fields in the theory through 
the interactions. Constraint equations for $\psi^{(-)}$ 
follow from the Euler-Lagrange equations,
\beq
   \label{eq:EL}
  \partial^\alpha\frac{\partial \cL}{\partial \partial^\alpha \psi}
  - \frac{\partial \cL}{\partial \psi} = 0 \, ,
\eeq
which can be written for every fermion field $\psi$ 
in the form
\beq
   \left(i \slash\partial - m_\psi\right)\psi \es
    i\partial^+\, \gamma^0\, \psi_I\,,
\eeq
where $\psi_I$ denotes the interaction terms.
Namely,
\beq
\label{cnui}
   {\nu_{i\,I}}
   \es -{1\over i\partial^+}
    U_{\mu i}^*\Gamma_\alpha\mu
    \left(g\, \bar n \Gamma_A^\alpha p-if\,\partial^\alpha\pi^\dag\right)\, ,\\
\label{cmu}
   {\mu_I}
   \es -{1\over i\partial^+}
    \Gamma_\alpha\nu_\mu \left(g\, \bar p \Gamma_g^\alpha n+if\,\partial^\alpha\pi\right)\, , \\
\label{cp}
   {p_I}
   \es -{1\over i\partial^+}
    g\Gamma^\alpha_A n \;\bar\nu_\mu\Gamma_\alpha \mu\, , \\
\label{cn}
   {n_I}
   \es -{1\over i\partial^+}
    g\Gamma^\alpha_A p \;\bar\mu\Gamma_\alpha \nu_\mu \, ,
\eeq
where $\Gamma_A^\alpha = \gamma^\alpha(1-g_A\gamma^5)$
and $\Gamma^\alpha = \gamma^\alpha(1-\gamma^5)$.
One can define a free field
\beq 
\label{psi0}        
    \psi_0     & = & \psi^{(-)}_0 + \psi^{(+)}_0   \,,
    \\\nnn
\eeq
where $\psi^{(-)}_0 = \psi^{(-)} - \psi^{(-)}_I$
and $\psi^{(+)}_0 \equiv \psi^{(+)}$ satisfy condition
\beq
   \psi^{(-)}_0 = {1\over i\partial^+} (i\alpha^\perp\partial^\perp
   +\beta m_\psi) \, \psi^{(+)}_0 \, .
\eeq
A calculation yields
\begin{align}
      {\partial \cL \over \partial \partial^- \psi } \partial^- \psi
   &- \bar\psi(i\slash\partial - m_\psi)\psi \\
    &=  \bar\psi_0 \gamma^+\frac{-(\partial^\perp)^2
  + m_\psi^2}{2i\partial^+}\psi_0
  - \bar\psi_I\Lambda^-(i\partial^+)\psi_I\, .\nnn
\end{align}
Since the interaction parts of fermion fields 
in on an effective theory with four-fermion interaction
terms satisfy mutually coupled constraint equations, 
they cannot be easily expressed in terms of the 
dynamically independent parts $\psi_0^{(+)}$. 
However, the weakness of electroweak interactions 
allows one to apply a weak-coupling expansion in 
solving the constraint equations approximately.
The interaction parts of all fields can be expanded 
in a series of powers of the coupling constants 
$g = G_F\cos\theta_C/\sqrt{2}$ and $f = F_\pi/\sqrt{2}$. 
The resulting FF Hamiltonian density $\cP^-$ 
takes the form of a series 
\begin{align}
\label{eq:P-}
\cP^- & =
   \cP^-_0 + \cP^-_1 + \cP^-_2 
   + o(g^k \, f^l:k+l \leq 2) \, , 
\end{align}
where
\begin{align}
\label{eq:P0}
\cP^-_0 & =
   \partial^\perp \pi^\dagger \partial^\perp\pi +
   m_\pi^2 \pi^\dagger \pi \nn
   &\quad + \sum_{\psi_0} \bar \psi_0 \gamma^+
   \frac{-(\partial^\perp)^2 + m_\psi^2}{ 2 i \partial^+} \psi_0 \, ,\\
\label{eq:P1}
\cP^-_1 & = 
   (if/2) \partial^+ \pi^\dagger J_{L0}^-
   - i f \partial^\perp \pi^\dagger J_{L0}^\perp
   - g J_{N0}^{\dagger\alpha} J_{L0\alpha} + H.c. \, ,
\end{align}
and $\cP^-_2$ denotes all terms of order $g^2$, $f^2$ and $gf$. 
The subscript $0$ in currents $J_N^\alpha$ and $J_L^\alpha$
indicates that the nucleon current $J_N^\alpha = \bar p 
\gamma^\alpha (1-g_A\gamma_5)n$ and lepton current 
$J_L^\alpha = \sum_i U_{\mu i}^*\bar \nu_i \gamma^\alpha
(1-\gamma_5)\mu$ are evaluated with all the free fermion
fields that are generically defined in Eq.~(\ref{psi0}).

It turns out that among all terms in $\cP^-_2$, only the 
terms proportional to $gf$ are important for calculation
in Sec.~\ref{FFOF}. These are the seagulls:
\beq
\label{Pgf}
\cP^-_{gf}
\es
   -igf \,\bar \mu_0
   \slash J_{N0} (1-\gamma^5) \frac{\gamma^+}{2i\partial^+} \, \slash\partial \pi^\dagger (1-\gamma^5)
   \mu_0 \nn
&& -igf \, \bar \nu_{\mu0} 
   \slash\partial\pi^\dagger(1-\gamma^5)  
   \frac{\gamma^+}{2i\partial^+} \, 
   \slash J_{N0}(1-\gamma^5)
   \nu_{\mu0} \nn
&& + H.c.\,.
\eeq
\bibliography{bibliography}

\begin{thebibliography}{49}
\expandafter\ifx\csname natexlab\endcsname\relax\def\natexlab#1{#1}\fi
\expandafter\ifx\csname bibnamefont\endcsname\relax
  \def\bibnamefont#1{#1}\fi
\expandafter\ifx\csname bibfnamefont\endcsname\relax
  \def\bibfnamefont#1{#1}\fi
\expandafter\ifx\csname citenamefont\endcsname\relax
  \def\citenamefont#1{#1}\fi
\expandafter\ifx\csname url\endcsname\relax
  \def\url#1{\texttt{#1}}\fi
\expandafter\ifx\csname urlprefix\endcsname\relax\def\urlprefix{URL }\fi
\providecommand{\bibinfo}[2]{#2}
\providecommand{\eprint}[2][]{\url{#2}}

\bibitem[{\citenamefont{Dirac}(1949)}]{Dirac:1949cp}
\bibinfo{author}{\bibfnamefont{P.~A.} \bibnamefont{Dirac}},
  \bibinfo{journal}{Rev.Mod.Phys.} \textbf{\bibinfo{volume}{21}},
  \bibinfo{pages}{392} (\bibinfo{year}{1949}).

\bibitem[{\citenamefont{Pontecorvo}(1968)}]{Pontecorvo:1967fh}
\bibinfo{author}{\bibfnamefont{B.}~\bibnamefont{Pontecorvo}},
  \bibinfo{journal}{Sov.Phys.JETP} \textbf{\bibinfo{volume}{26}},
  \bibinfo{pages}{984} (\bibinfo{year}{1968}).

\bibitem[{\citenamefont{Bilenky and Pontecorvo}(1977)}]{Bilenky:1977ne}
\bibinfo{author}{\bibfnamefont{S.~M.} \bibnamefont{Bilenky}} \bibnamefont{and}
  \bibinfo{author}{\bibfnamefont{B.}~\bibnamefont{Pontecorvo}},
  \bibinfo{journal}{Comments Nucl. Part. Phys.} \textbf{\bibinfo{volume}{7}},
  \bibinfo{pages}{149} (\bibinfo{year}{1977}).

\bibitem[{\citenamefont{Kayser}(1981)}]{Kayser:1981ye}
\bibinfo{author}{\bibfnamefont{B.}~\bibnamefont{Kayser}},
  \bibinfo{journal}{Phys. Rev.} \textbf{\bibinfo{volume}{D24}},
  \bibinfo{pages}{110} (\bibinfo{year}{1981}).

\bibitem[{\citenamefont{Rich}(1993)}]{Rich:1993wu}
\bibinfo{author}{\bibfnamefont{J.}~\bibnamefont{Rich}}, \bibinfo{journal}{Phys.
  Rev.} \textbf{\bibinfo{volume}{D48}}, \bibinfo{pages}{4318}
  (\bibinfo{year}{1993}).

\bibitem[{\citenamefont{Giunti et~al.}(1991)\citenamefont{Giunti, Kim, and
  Lee}}]{Giunti:1991ca}
\bibinfo{author}{\bibfnamefont{C.}~\bibnamefont{Giunti}},
  \bibinfo{author}{\bibfnamefont{C.}~\bibnamefont{Kim}}, \bibnamefont{and}
  \bibinfo{author}{\bibfnamefont{U.}~\bibnamefont{Lee}},
  \bibinfo{journal}{Phys.Rev.} \textbf{\bibinfo{volume}{D44}},
  \bibinfo{pages}{3635} (\bibinfo{year}{1991}).

\bibitem[{\citenamefont{Giunti et~al.}(1993)\citenamefont{Giunti, Kim, Lee, and
  Lee}}]{Giunti:1993se}
\bibinfo{author}{\bibfnamefont{C.}~\bibnamefont{Giunti}},
  \bibinfo{author}{\bibfnamefont{C.}~\bibnamefont{Kim}},
  \bibinfo{author}{\bibfnamefont{J.}~\bibnamefont{Lee}}, \bibnamefont{and}
  \bibinfo{author}{\bibfnamefont{U.}~\bibnamefont{Lee}},
  \bibinfo{journal}{Phys.Rev.} \textbf{\bibinfo{volume}{D48}},
  \bibinfo{pages}{4310} (\bibinfo{year}{1993}), \eprint{hep-ph/9305276}.

\bibitem[{\citenamefont{Grimus and Stockinger}(1996)}]{Grimus:1996av}
\bibinfo{author}{\bibfnamefont{W.}~\bibnamefont{Grimus}} \bibnamefont{and}
  \bibinfo{author}{\bibfnamefont{P.}~\bibnamefont{Stockinger}},
  \bibinfo{journal}{Phys.Rev.} \textbf{\bibinfo{volume}{D54}},
  \bibinfo{pages}{3414} (\bibinfo{year}{1996}), \eprint{hep-ph/9603430}.

\bibitem[{\citenamefont{Beuthe}(2003)}]{Beuthe:2001rc}
\bibinfo{author}{\bibfnamefont{M.}~\bibnamefont{Beuthe}},
  \bibinfo{journal}{Phys.Rept.} \textbf{\bibinfo{volume}{375}},
  \bibinfo{pages}{105} (\bibinfo{year}{2003}), \eprint{hep-ph/0109119}.

\bibitem[{\citenamefont{Giunti}(2002)}]{Giunti:2002xg}
\bibinfo{author}{\bibfnamefont{C.}~\bibnamefont{Giunti}},
  \bibinfo{journal}{JHEP} \textbf{\bibinfo{volume}{0211}}, \bibinfo{pages}{017}
  (\bibinfo{year}{2002}), \eprint{hep-ph/0205014}.

\bibitem[{\citenamefont{Giunti}(2004)}]{Giunti:2003ax}
\bibinfo{author}{\bibfnamefont{C.}~\bibnamefont{Giunti}},
  \bibinfo{journal}{Found.Phys.Lett.} \textbf{\bibinfo{volume}{17}},
  \bibinfo{pages}{103} (\bibinfo{year}{2004}), \eprint{hep-ph/0302026}.

\bibitem[{\citenamefont{Cohen et~al.}(2009)\citenamefont{Cohen, Glashow, and
  Ligeti}}]{Cohen:2008qb}
\bibinfo{author}{\bibfnamefont{A.~G.} \bibnamefont{Cohen}},
  \bibinfo{author}{\bibfnamefont{S.~L.} \bibnamefont{Glashow}},
  \bibnamefont{and} \bibinfo{author}{\bibfnamefont{Z.}~\bibnamefont{Ligeti}},
  \bibinfo{journal}{Phys.Lett.} \textbf{\bibinfo{volume}{B678}},
  \bibinfo{pages}{191} (\bibinfo{year}{2009}), \eprint{0810.4602}.

\bibitem[{\citenamefont{Akhmedov and Smirnov}(2009)}]{Akhmedov:2009rb}
\bibinfo{author}{\bibfnamefont{E.~K.} \bibnamefont{Akhmedov}} \bibnamefont{and}
  \bibinfo{author}{\bibfnamefont{A.~Y.} \bibnamefont{Smirnov}},
  \bibinfo{journal}{Phys.Atom.Nucl.} \textbf{\bibinfo{volume}{72}},
  \bibinfo{pages}{1363} (\bibinfo{year}{2009}), \eprint{0905.1903}.

\bibitem[{\citenamefont{Merle}(2009)}]{Merle:2009re}
\bibinfo{author}{\bibfnamefont{A.}~\bibnamefont{Merle}},
  \bibinfo{journal}{Phys.Rev.} \textbf{\bibinfo{volume}{C80}},
  \bibinfo{pages}{054616} (\bibinfo{year}{2009}), \eprint{0907.3554}.

\bibitem[{\citenamefont{Akhmedov and Kopp}(2010)}]{Akhmedov:2010ms}
\bibinfo{author}{\bibfnamefont{E.~K.} \bibnamefont{Akhmedov}} \bibnamefont{and}
  \bibinfo{author}{\bibfnamefont{J.}~\bibnamefont{Kopp}},
  \bibinfo{journal}{JHEP} \textbf{\bibinfo{volume}{1004}}, \bibinfo{pages}{008}
  (\bibinfo{year}{2010}), \eprint{1001.4815}.

\bibitem[{\citenamefont{Fukuda et~al.}(1998)}]{Fukuda:1998mi}
\bibinfo{author}{\bibfnamefont{Y.}~\bibnamefont{Fukuda}} \bibnamefont{et~al.}
  (\bibinfo{collaboration}{Super-Kamiokande Collaboration}),
  \bibinfo{journal}{Phys.Rev.Lett.} \textbf{\bibinfo{volume}{81}},
  \bibinfo{pages}{1562} (\bibinfo{year}{1998}), \eprint{hep-ex/9807003}.

\bibitem[{\citenamefont{Ahmad et~al.}(2002)}]{Ahmad:2002jz}
\bibinfo{author}{\bibfnamefont{Q.}~\bibnamefont{Ahmad}} \bibnamefont{et~al.}
  (\bibinfo{collaboration}{SNO Collaboration}),
  \bibinfo{journal}{Phys.Rev.Lett.} \textbf{\bibinfo{volume}{89}},
  \bibinfo{pages}{011301} (\bibinfo{year}{2002}), \eprint{nucl-ex/0204008}.

\bibitem[{\citenamefont{Eguchi et~al.}(2003)}]{Eguchi:2002dm}
\bibinfo{author}{\bibfnamefont{K.}~\bibnamefont{Eguchi}} \bibnamefont{et~al.}
  (\bibinfo{collaboration}{KamLAND Collaboration}),
  \bibinfo{journal}{Phys.Rev.Lett.} \textbf{\bibinfo{volume}{90}},
  \bibinfo{pages}{021802} (\bibinfo{year}{2003}), \eprint{hep-ex/0212021}.

\bibitem[{\citenamefont{Abe et~al.}(2012)}]{Abe:2012gx}
\bibinfo{author}{\bibfnamefont{K.}~\bibnamefont{Abe}} \bibnamefont{et~al.}
  (\bibinfo{collaboration}{T2K Collaboration}), \bibinfo{journal}{Phys.Rev.}
  \textbf{\bibinfo{volume}{D85}}, \bibinfo{pages}{031103(R)}
  (\bibinfo{year}{2012}), \eprint{1201.1386}.

\bibitem[{\citenamefont{Adamson et~al.}(2012)}]{Adamson:2012rm}
\bibinfo{author}{\bibfnamefont{P.}~\bibnamefont{Adamson}} \bibnamefont{et~al.}
  (\bibinfo{collaboration}{MINOS Collaboration}),
  \bibinfo{journal}{Phys.Rev.Lett.} \textbf{\bibinfo{volume}{108}},
  \bibinfo{pages}{191801} (\bibinfo{year}{2012}), \eprint{1202.2772}.

\bibitem[{\citenamefont{Gell-Mann and Goldberger}(1953)}]{GellMann:1953zz}
\bibinfo{author}{\bibfnamefont{M.}~\bibnamefont{Gell-Mann}} \bibnamefont{and}
  \bibinfo{author}{\bibfnamefont{M.~L.} \bibnamefont{Goldberger}},
  \bibinfo{journal}{Phys. Rev.} \textbf{\bibinfo{volume}{91}},
  \bibinfo{pages}{398} (\bibinfo{year}{1953}).

\bibitem[{\citenamefont{G{\l}azek and Trawi{\'n}ski}(2012)}]{Glazek:2012pd}
\bibinfo{author}{\bibfnamefont{S.~D.} \bibnamefont{G{\l}azek}}
  \bibnamefont{and} \bibinfo{author}{\bibfnamefont{A.~P.}
  \bibnamefont{Trawi{\'n}ski}}, \bibinfo{journal}{Phys.Rev.}
  \textbf{\bibinfo{volume}{D85}}, \bibinfo{pages}{125001}
  (\bibinfo{year}{2012}), \eprint{1204.6007}.

\bibitem[{\citenamefont{Feynman and Gell-Mann}(1958)}]{Feynman:1958ty}
\bibinfo{author}{\bibfnamefont{R.}~\bibnamefont{Feynman}} \bibnamefont{and}
  \bibinfo{author}{\bibfnamefont{M.}~\bibnamefont{Gell-Mann}},
  \bibinfo{journal}{Phys.Rev.} \textbf{\bibinfo{volume}{109}},
  \bibinfo{pages}{193} (\bibinfo{year}{1958}).

\bibitem[{\citenamefont{Gell-Mann and Levy}(1960)}]{GellMann:1960np}
\bibinfo{author}{\bibfnamefont{M.}~\bibnamefont{Gell-Mann}} \bibnamefont{and}
  \bibinfo{author}{\bibfnamefont{M.}~\bibnamefont{Levy}},
  \bibinfo{journal}{Nuovo Cim.} \textbf{\bibinfo{volume}{16}},
  \bibinfo{pages}{705} (\bibinfo{year}{1960}).

\bibitem[{\citenamefont{Dirac}(1965)}]{PhysRev.139.B684}
\bibinfo{author}{\bibfnamefont{P.~A.~M.} \bibnamefont{Dirac}},
  \bibinfo{journal}{Phys. Rev.} \textbf{\bibinfo{volume}{139}},
  \bibinfo{pages}{B684} (\bibinfo{year}{1965}).

\bibitem[{\citenamefont{Nambu and Jona-Lasinio}(1961)}]{Nambu:1961tp}
\bibinfo{author}{\bibfnamefont{Y.}~\bibnamefont{Nambu}} \bibnamefont{and}
  \bibinfo{author}{\bibfnamefont{G.}~\bibnamefont{Jona-Lasinio}},
  \bibinfo{journal}{Phys.Rev.} \textbf{\bibinfo{volume}{122}},
  \bibinfo{pages}{345} (\bibinfo{year}{1961}).

\bibitem[{\citenamefont{Kogut and Susskind}(1973)}]{Kogut:1972di}
\bibinfo{author}{\bibfnamefont{J.~B.} \bibnamefont{Kogut}} \bibnamefont{and}
  \bibinfo{author}{\bibfnamefont{L.}~\bibnamefont{Susskind}},
  \bibinfo{journal}{Phys.Rept.} \textbf{\bibinfo{volume}{8}},
  \bibinfo{pages}{75} (\bibinfo{year}{1973}).

\bibitem[{\citenamefont{Leutwyler and Stern}(1978)}]{Leutwyler:1977vy}
\bibinfo{author}{\bibfnamefont{H.}~\bibnamefont{Leutwyler}} \bibnamefont{and}
  \bibinfo{author}{\bibfnamefont{J.}~\bibnamefont{Stern}},
  \bibinfo{journal}{Annals Phys.} \textbf{\bibinfo{volume}{112}},
  \bibinfo{pages}{94} (\bibinfo{year}{1978}).

\bibitem[{\citenamefont{Shifman et~al.}(1979)\citenamefont{Shifman, Vainshtein,
  and Zakharov}}]{Shifman:1978bx}
\bibinfo{author}{\bibfnamefont{M.~A.} \bibnamefont{Shifman}},
  \bibinfo{author}{\bibfnamefont{A.}~\bibnamefont{Vainshtein}},
  \bibnamefont{and} \bibinfo{author}{\bibfnamefont{V.~I.}
  \bibnamefont{Zakharov}}, \bibinfo{journal}{Nucl.Phys.}
  \textbf{\bibinfo{volume}{B147}}, \bibinfo{pages}{385} (\bibinfo{year}{1979}).

\bibitem[{\citenamefont{Gasser and Leutwyler}(1984)}]{Gasser:1983yg}
\bibinfo{author}{\bibfnamefont{J.}~\bibnamefont{Gasser}} \bibnamefont{and}
  \bibinfo{author}{\bibfnamefont{H.}~\bibnamefont{Leutwyler}},
  \bibinfo{journal}{Annals Phys.} \textbf{\bibinfo{volume}{158}},
  \bibinfo{pages}{142} (\bibinfo{year}{1984}).

\bibitem[{\citenamefont{Weinberg}(1989)}]{Weinberg:1988cp}
\bibinfo{author}{\bibfnamefont{S.}~\bibnamefont{Weinberg}},
  \bibinfo{journal}{Rev.Mod.Phys.} \textbf{\bibinfo{volume}{61}},
  \bibinfo{pages}{1} (\bibinfo{year}{1989}).

\bibitem[{\citenamefont{Wilson et~al.}(1994)}]{Wilson:1994fk}
\bibinfo{author}{\bibfnamefont{K.~G.} \bibnamefont{Wilson}}
  \bibnamefont{et~al.}, \bibinfo{journal}{Phys.Rev.}
  \textbf{\bibinfo{volume}{D49}}, \bibinfo{pages}{6720} (\bibinfo{year}{1994}),
  \eprint{hep-th/9401153}.

\bibitem[{\citenamefont{Brodsky and Shrock}(2011)}]{Brodsky:2009zd}
\bibinfo{author}{\bibfnamefont{S.~J.} \bibnamefont{Brodsky}} \bibnamefont{and}
  \bibinfo{author}{\bibfnamefont{R.}~\bibnamefont{Shrock}},
  \bibinfo{journal}{Proc.Nat.Acad.Sci.} \textbf{\bibinfo{volume}{108}},
  \bibinfo{pages}{45} (\bibinfo{year}{2011}), \eprint{0905.1151}.

\bibitem[{\citenamefont{Weinberg}(2011)}]{Weinberg:2010wq}
\bibinfo{author}{\bibfnamefont{S.}~\bibnamefont{Weinberg}},
  \bibinfo{journal}{Phys.Rev.} \textbf{\bibinfo{volume}{D83}},
  \bibinfo{pages}{063508} (\bibinfo{year}{2011}), \eprint{1011.1630}.

\bibitem[{\citenamefont{Brodsky et~al.}(2012)\citenamefont{Brodsky, Roberts,
  Shrock, and Tandy}}]{Brodsky:2012ku}
\bibinfo{author}{\bibfnamefont{S.~J.} \bibnamefont{Brodsky}},
  \bibinfo{author}{\bibfnamefont{C.~D.} \bibnamefont{Roberts}},
  \bibinfo{author}{\bibfnamefont{R.}~\bibnamefont{Shrock}}, \bibnamefont{and}
  \bibinfo{author}{\bibfnamefont{P.~C.} \bibnamefont{Tandy}}
  (\bibinfo{year}{2012}), \eprint{1202.2376}.

\bibitem[{\citenamefont{G{\l}azek}(2012)}]{PhysRevD.85.125018}
\bibinfo{author}{\bibfnamefont{S.~D.} \bibnamefont{G{\l}azek}},
  \bibinfo{journal}{Phys. Rev. D} \textbf{\bibinfo{volume}{85}},
  \bibinfo{pages}{125018} (\bibinfo{year}{2012}).

\bibitem[{\citenamefont{{'t Hooft} and Veltman}(1972)}]{'tHooft:1972fi}
\bibinfo{author}{\bibfnamefont{G.}~\bibnamefont{{'t Hooft}}} \bibnamefont{and}
  \bibinfo{author}{\bibfnamefont{M.}~\bibnamefont{Veltman}},
  \bibinfo{journal}{Nucl.Phys.} \textbf{\bibinfo{volume}{B44}},
  \bibinfo{pages}{189} (\bibinfo{year}{1972}).

\bibitem[{\citenamefont{Akhmedov et~al.}(2012)\citenamefont{Akhmedov,
  Hernandez, and Smirnov}}]{Akhmedov:2012uu}
\bibinfo{author}{\bibfnamefont{E.}~\bibnamefont{Akhmedov}},
  \bibinfo{author}{\bibfnamefont{D.}~\bibnamefont{Hernandez}},
  \bibnamefont{and} \bibinfo{author}{\bibfnamefont{A.}~\bibnamefont{Smirnov}},
  \bibinfo{journal}{JHEP} \textbf{\bibinfo{volume}{1204}}, \bibinfo{pages}{052}
  (\bibinfo{year}{2012}), \eprint{1201.4128}.

\bibitem[{\citenamefont{Brodsky et~al.}(1973)\citenamefont{Brodsky, Roskies,
  and Suaya}}]{Brodsky:1973kb}
\bibinfo{author}{\bibfnamefont{S.~J.} \bibnamefont{Brodsky}},
  \bibinfo{author}{\bibfnamefont{R.}~\bibnamefont{Roskies}}, \bibnamefont{and}
  \bibinfo{author}{\bibfnamefont{R.}~\bibnamefont{Suaya}},
  \bibinfo{journal}{Phys.Rev.} \textbf{\bibinfo{volume}{D8}},
  \bibinfo{pages}{4574} (\bibinfo{year}{1973}).

\bibitem[{\citenamefont{Kogut and Soper}(1970)}]{Kogut:1969xa}
\bibinfo{author}{\bibfnamefont{J.~B.} \bibnamefont{Kogut}} \bibnamefont{and}
  \bibinfo{author}{\bibfnamefont{D.~E.} \bibnamefont{Soper}},
  \bibinfo{journal}{Phys.Rev.} \textbf{\bibinfo{volume}{D1}},
  \bibinfo{pages}{2901} (\bibinfo{year}{1970}).

\bibitem[{\citenamefont{Bjorken et~al.}(1971)\citenamefont{Bjorken, Kogut, and
  Soper}}]{Bjorken:1970ah}
\bibinfo{author}{\bibfnamefont{J.}~\bibnamefont{Bjorken}},
  \bibinfo{author}{\bibfnamefont{J.~B.} \bibnamefont{Kogut}}, \bibnamefont{and}
  \bibinfo{author}{\bibfnamefont{D.~E.} \bibnamefont{Soper}},
  \bibinfo{journal}{Phys.Rev.} \textbf{\bibinfo{volume}{D3}},
  \bibinfo{pages}{1382} (\bibinfo{year}{1971}).

\bibitem[{\citenamefont{Neville and
  Rohrlich}(1971{\natexlab{a}})}]{Neville:1971zk}
\bibinfo{author}{\bibfnamefont{R.}~\bibnamefont{Neville}} \bibnamefont{and}
  \bibinfo{author}{\bibfnamefont{F.}~\bibnamefont{Rohrlich}},
  \bibinfo{journal}{Nuovo Cim.} \textbf{\bibinfo{volume}{A1}},
  \bibinfo{pages}{625} (\bibinfo{year}{1971}{\natexlab{a}}).

\bibitem[{\citenamefont{Neville and
  Rohrlich}(1971{\natexlab{b}})}]{Neville:1971uc}
\bibinfo{author}{\bibfnamefont{R.}~\bibnamefont{Neville}} \bibnamefont{and}
  \bibinfo{author}{\bibfnamefont{F.}~\bibnamefont{Rohrlich}},
  \bibinfo{journal}{Phys.Rev.} \textbf{\bibinfo{volume}{D3}},
  \bibinfo{pages}{1692} (\bibinfo{year}{1971}{\natexlab{b}}).

\bibitem[{\citenamefont{Chang et~al.}(1973)\citenamefont{Chang, Root, and
  Yan}}]{Chang:1972xt}
\bibinfo{author}{\bibfnamefont{S.-J.} \bibnamefont{Chang}},
  \bibinfo{author}{\bibfnamefont{R.~G.} \bibnamefont{Root}}, \bibnamefont{and}
  \bibinfo{author}{\bibfnamefont{T.-M.} \bibnamefont{Yan}},
  \bibinfo{journal}{Phys.Rev.} \textbf{\bibinfo{volume}{D7}},
  \bibinfo{pages}{1133} (\bibinfo{year}{1973}).

\bibitem[{\citenamefont{Chang and Yan}(1973)}]{Chang:1973qi}
\bibinfo{author}{\bibfnamefont{S.-J.} \bibnamefont{Chang}} \bibnamefont{and}
  \bibinfo{author}{\bibfnamefont{T.-M.} \bibnamefont{Yan}},
  \bibinfo{journal}{Phys.Rev.} \textbf{\bibinfo{volume}{D7}},
  \bibinfo{pages}{1147} (\bibinfo{year}{1973}).

\bibitem[{\citenamefont{Yan}(1973{\natexlab{a}})}]{Yan:1973qf}
\bibinfo{author}{\bibfnamefont{T.-M.} \bibnamefont{Yan}},
  \bibinfo{journal}{Phys.Rev.} \textbf{\bibinfo{volume}{D7}},
  \bibinfo{pages}{1760} (\bibinfo{year}{1973}{\natexlab{a}}).

\bibitem[{\citenamefont{Yan}(1973{\natexlab{b}})}]{Yan:1973qg}
\bibinfo{author}{\bibfnamefont{T.-M.} \bibnamefont{Yan}},
  \bibinfo{journal}{Phys.Rev.} \textbf{\bibinfo{volume}{D7}},
  \bibinfo{pages}{1780} (\bibinfo{year}{1973}{\natexlab{b}}).

\bibitem[{\citenamefont{Ten~Eyck and Rohrlich}(1974)}]{TenEyck:1974cy}
\bibinfo{author}{\bibfnamefont{J.~H.} \bibnamefont{Ten~Eyck}} \bibnamefont{and}
  \bibinfo{author}{\bibfnamefont{F.}~\bibnamefont{Rohrlich}},
  \bibinfo{journal}{Phys.Rev.} \textbf{\bibinfo{volume}{D9}},
  \bibinfo{pages}{2237} (\bibinfo{year}{1974}).

\bibitem[{\citenamefont{Brodsky et~al.}(1998)\citenamefont{Brodsky, Pauli, and
  Pinsky}}]{Brodsky:1997de}
\bibinfo{author}{\bibfnamefont{S.~J.} \bibnamefont{Brodsky}},
  \bibinfo{author}{\bibfnamefont{H.-C.} \bibnamefont{Pauli}}, \bibnamefont{and}
  \bibinfo{author}{\bibfnamefont{S.~S.} \bibnamefont{Pinsky}},
  \bibinfo{journal}{Phys.Rept.} \textbf{\bibinfo{volume}{301}},
  \bibinfo{pages}{299} (\bibinfo{year}{1998}), \eprint{hep-ph/9705477}.

\end{thebibliography}
\end{document}